\documentstyle[epsfig,longtable]{aipproc}

\begin{document}
\begin{flushright}
Cavendish-HEP-97/07 \\
hep-ph/9706263
\end{flushright}
\title{Power Corrections in Deep Inelastic Event Shape Variables}
\author{Mrinal Dasgupta\thanks{Talk given at the 5th International
Workshop on Deep Inelastic Scattering, Chicago, April 1997.}}
\address{Cavendish Laboratory,University of Cambridge,Madingley
Road,Cambridge,CB3 0HE,United Kingdom}

\maketitle

\begin{abstract}
Results are presented for the leading power-suppressed ($1/Q$) contributions
to shape variables in deep inelastic lepton scattering. These are then
combined with available leading order perturbative estimates.
\end{abstract}

\section*{Introduction}
The increasing quantity of data from HERA has given a great boost to
the testing of QCD via determinations of the strong coupling constant
and the possibility of observing, over a wide kinematic range, its
running behaviour.
\\
However in any comparison between theory and experimental data
proper account has to be taken of the non-perturbative behaviour
inherent in any observable QCD process. 
While event shape variables are considered useful observables for
$\alpha_{s}$ determinations they can suffer from significant power
suppressed effects which are non-perturbative in origin.
These $1/Q$ corrections have already been calculated for a number of
$e^{+}e^-$ event shapes [1] and similar results are obtained here for DIS.
To make the analogy with $e^{+}e^-$ shape variables clearer it proves
useful to work in the Breit frame and to define current jet event
shapes as done in the next section.
The dispersive approach [2] is used to estimate the power corrections,
which is equivalent to a calculation of the leading renormalon ambiguity.
\section*{Event shape variables}
One can define several infrared safe quantities that characterize the
shape of the event in the current hemisphere $H_c$ of the Breit
frame. 
 Perturbative results are available in [3]. The event shapes considered below are the
thrust, the jet mass, the jet broadening and the $C$ parameter with the
following definitions in the above order where $\vec{n}$ denotes a unit
3-vector along the current direction and the summation index $a$ runs
over only those partons that are in the current hemisphere.
\begin{equation}
T_{Q}=\frac{\sum_{a}\vec{p}_{a}.\vec{n}}{Q/2}
\end{equation}
\begin{equation}
\rho_{Q}=\frac{({\sum_{a}{p}_{a}})^{2}}{Q^{2}}
\end{equation}
\begin{equation} 
B_{Q} = \frac{\sum_{a}|\vec{p}_{a}\times \vec{n}|}{Q/2}
\end{equation}
\begin{equation}
C_{Q}=3(\lambda_{1}\lambda_{2} + \lambda_{2}\lambda_{3} +
\lambda_{3}\lambda_{1})
\end{equation}
where the $\lambda_{i}$ denote the eigenvalues of the symmetric matrix
$\Theta^{ij}$ defined as
\begin{equation}
\Theta ^{ij} = 2\sum_{a} \left (p_{a}^{i}p_{a}^{j}/|p_{a}| \right )/Q
\end{equation}
The subscript $Q$ indicates the normalization factor $Q/2$. 
One can also use a normalization factor given by the sum of
energies of partons in the Breit frame, but the
normalization does not affect the power corrections we compute.
\section*{Dispersive Method}
Our estimate of the leading power corrections to the perturbative
results are based on the approach of
Ref.[2]. Non-perturbative effects at long distances are
assumed to give rise to a modification $\delta\alpha_{eff}(\mu^{2})$
in the QCD effective coupling at low values of the scale
$\mu^{2}$. The effect on some observable $F$ is then given by a
characteristic function ${\mathcal{F}} (x,\epsilon)$ as follows:
\begin{equation}
\delta F(x,Q^2) = \int_{0}^{\infty} \frac{d\mu^2}{\mu^{2}}
\delta\alpha_{eff}(\mu^2) \dot{\mathcal{F}}(x,\epsilon=\mu^2/Q^2)
\end{equation}
where 
\begin{equation}
\dot{\mathcal{F}}(x,\epsilon) = -\epsilon \frac{\partial}{\partial
\epsilon}{\mathcal{F}}(x,\epsilon)
\end{equation}
The characteristic function is obtained by computing the relevant
one-loop graphs with a non-zero gluon mass $\mu$ [4].
\\
Arbitrary finite modifications of the effective coupling at low scales
would generally introduce power corrections of the form $1/k^{2p}$ into
the ultraviolet behaviour of the running coupling itself.
As discussed in Ref.[2], such a modification would
destroy the basis of the operator product expansion [5].
This leads to the constraint that only terms in the small-$\epsilon$
behaviour of the characteristic function that are non-analytic at
$\epsilon=0$ will lead to power behaved non-perturbative contributions
[2].
\section*{Results}
We find that in DIS as for $e^{+}e^{-}$ the leading non-analytic
behaviour takes the form of $\sqrt{\epsilon}$ with an extra $\ln{\epsilon}$ in the
jet broadening. The details of the calculation can be found in
[3].
The results for the power corrections are presented below. $A_1$ is an
phenomenological parameter first suggested in Ref.[2] which is
expected to be approximately universal. $e^+e^-$ data are consistent with a value
for $A_{1}$ of about $0.25$ GeV.
In the case of the current jet thrust we get a non-perturbative
contribution
\begin{equation}
\delta \langle 1-T_Q \rangle = 4 \frac{A_1}{Q}
\end{equation}
\begin{figure}
\begin{center}
\epsfig{file=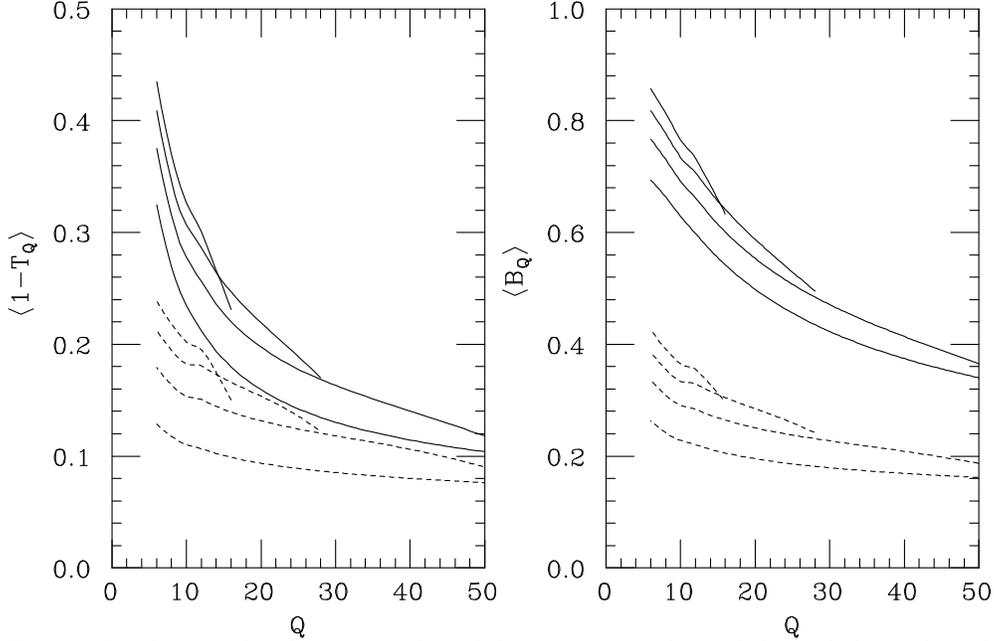,height=8.0cm}
\caption{Predictions for the mean value of the current jet
thrust(left) and the jet broadening
in deep inelastic scattering at $\sqrt S=296$ GeV. The 4 plots shown
in each case are for $x$ =0.003, 0.01, 0.03,and 0.1. The dashed curves
are the leading order perturbative prediction while the solid curves
represent the combined leading order perturbative and power behaved contributions} 
\end{center}
\end{figure} 

For the current jet mass one gets a correction given by 
\begin{equation}
\delta \langle \rho_{Q} \rangle = 2 \frac{A_1}{Q}
\end{equation}
\begin{figure}
\begin{center}
\epsfig{file=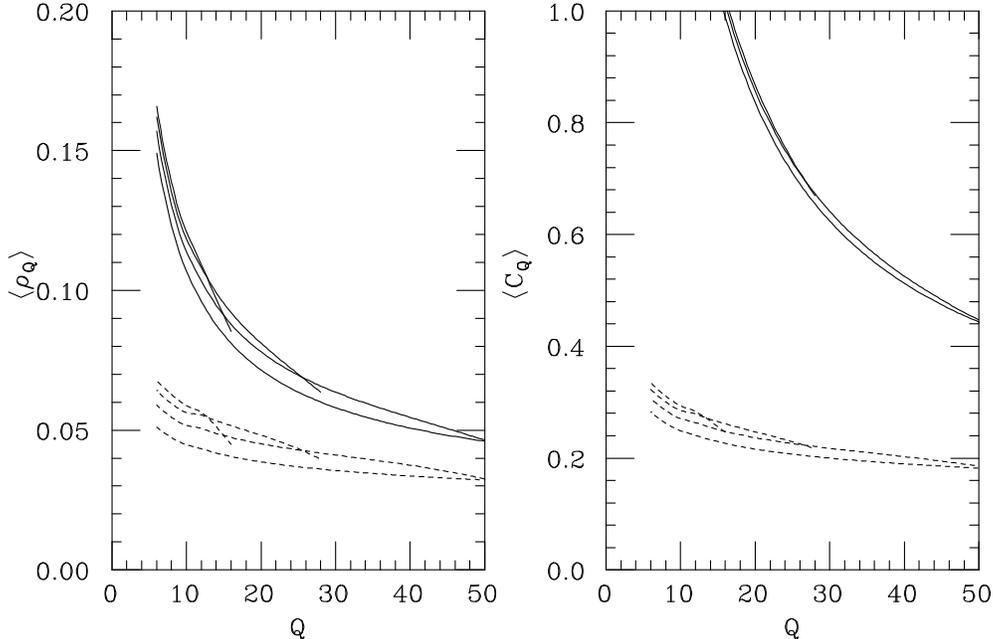,height=8.0cm}
\caption{Predictions for the mean value of the current jet mass(left) and $C$ parameter
in deep inelastic scattering at $\sqrt S=296$ GeV. Curves as in Fig.1}
\end{center}
\end{figure}

For the $C$ parameter we get
\begin{equation}
\delta \langle C_Q \rangle = 12 \pi \frac{A_1}{Q}
\end{equation}

In the case of the jet broadening a slightly different behaviour is
obtained 
\begin{equation}
\delta \langle B_{Q} \rangle = 8\frac{A_{1}}{Q} \ln{\frac{Q}{Q_0}}
\end{equation}
where $Q_0$ is an undetermined scale.
The plots for the
combined leading order prediction and power correction are presented
for all of these event shape variables in figures 1-2. 
The MRS$A^{\prime}$ [8] parton distribution was used in every case. 
\section*{Conclusions}
It should be pointed out that as explained in Ref.[3] the
predictions for  the numerical coefficients of  $\frac{A_1}{Q}$ for the $C$
parameter and the jet broadening are somewhat uncertain. This is due
to an inherent limitation of the massive gluon technique with regard
to event shapes. The use of a gluon mass in perturbative calculations
is equivalent to summing inclusively over all bubbles in the
renormalon chain. In the case of less inclusive variables like event
shapes the phase space is dependent on the bubble structure which is
ignored by the massive gluon formulation. This has been shown for
event shapes in [6] and for gluon fragmentation functions
in Ref.[7]. It is believed that this limitation does not affect
the form of the power corrections to any of the shape variables and in fact appears not to make any
significant difference to the predictions for the thrust and the jet
mass where our results are not sensitive to the inclusion of a gluon
mass in the definition of the shape variable. The power corrections
here arise from using a massive gluon phase space alone. In the case of the
$C$ parameter and the jet broadening the predictions for the numerical
coefficients depend on whether or not a gluon mass is included
throughout.
A glance at the relevant figures also reveals that there are
enormously large power corrections in the case of $C_Q$ and $B_Q$ even at
$Q=50$ GeV.
This fact
coupled with the uncertainty about the coefficient would mean that
these variables are unsuitable for an extraction of the strong
coupling.
On the other hand the predictions for power corrections to the thrust
and the jet mass are more reliable making these, in our opinion, good
observables for determination of $\alpha_s$.
The relevant experimental results have been presented at this workshop
by the H1 collaboration [9,10]

\section*{Acknowledgements}
All of the work presented here was done in collaboration with
B.R.Webber.
I am also grateful to Trinity College, Cambridge and the Committee of
Vice-Chancellors and Principals of the Universities of the United
Kingdom for financial support.

\end{document}